\documentclass[runningheads, a4paper]{llncs}
\usepackage{amssymb, amsmath, bm, latexsym, comment}
\usepackage{multirow}
\usepackage{xcolor}
\usepackage{graphicx}
\usepackage{subfigure}
\usepackage{indentfirst}
\usepackage{setspace}
\usepackage{verbatim}
\usepackage{array}
\usepackage{cite}
\usepackage{booktabs}
\usepackage{hyperref}
\usepackage[capitalize]{cleveref}
\usepackage{arydshln}
\usepackage{multirow}

\newcolumntype{L}[1]{>{\raggedright\let\newline\\\arraybackslash\hspace{0pt}}m{#1}}
\newcolumntype{C}[1]{>{\centering\let\newline\\\arraybackslash\hspace{0pt}}m{#1}}
\newcolumntype{R}[1]{>{\raggedleft\let\newline\\\arraybackslash\hspace{0pt}}m{#1}}
\def\ourmodel{DVasMesh}
\begin{document}

\title{DVasMesh: Deep Structured Mesh Reconstruction from Vascular Images for Dynamics Modeling of Vessels}

\titlerunning{DVasMesh}

\author{Dengqiang Jia\inst{1}, Xinnian Yang\inst{2, 3}, Xiaosong Xiong\inst{3}, Shijie Huang\inst{3},  Feiyu Hou\inst{4}, Li Qin\inst{\rm 5}, Kaicong Sun \inst{3},  Kannie Wai Yan Chan\inst{1,6}, Dinggang Shen\inst{3,7,8}
\thanks{Corresponding author: Dinggang.Shen@gmail.com.}
}

\authorrunning{Dengqiang Jia et al.}

\tocauthor{}

\institute{
\textsuperscript{\rm 1}Hong Kong Centre for Cerebro-cardiovascular Health Engineering, Hong Kong 000000, China\\
\email{dqjia@hkcoche.org}\\
 \textsuperscript{\rm 2}School of Data Science, City University of Hong Kong, Hong Kong 000000, China\\
  \textsuperscript{\rm 3}School of Biomedical Engineering \& State Key Laboratory of Advanced Medical Materials and Devices, ShanghaiTech University, Shanghai 201210, China \\
   \textsuperscript{\rm 4}School of Mechanical and Automotive Engineering, Shanghai University of Engineering Science, Shanghai 201620, China\\
   \email{Dinggang.Shen@gmail.com}\\
   \textsuperscript{\rm 5}JinFeng Laboratory, Chongqing 401329, China\\
    \textsuperscript{\rm 6}Department of Biomedical Engineering, City University of Hong Kong, Hong Kong 000000, China\\
   \textsuperscript{\rm 7}Shanghai United Imaging Intelligence Co., Ltd., Shanghai 200230, China\\
   \textsuperscript{\rm 8}Shanghai Clinical Research and Trial Center, Shanghai 201210, China \\
  }

\maketitle

\begin{abstract}
Vessel dynamics simulation
is vital in studying the relationship between geometry and vascular disease progression. 
Reliable dynamics simulation relies on high-quality vascular meshes. Most of the existing mesh generation methods highly depend on manual annotation, which is time-consuming and laborious, 
usually facing challenges such as branch merging and vessel disconnection. 
This will hinder vessel dynamics simulation, especially for the population study.
To address this issue, we propose a deep learning-based method, dubbed as \ourmodel{} to directly generate structured hexahedral vascular meshes from vascular images. Our contributions are threefold. First, we propose to formally formulate each vertex of the vascular graph by a four-element vector, including coordinates of the centerline point and the radius. 
Second, a vectorized graph template is employed to guide \ourmodel{} to estimate the vascular graph. Specifically, we introduce a sampling operator, which samples the extracted features of the vascular image (by a segmentation network) according to the vertices in the template graph. Third, we employ a graph convolution network (GCN) and take the sampled features as nodes to estimate the deformation between vertices of the template graph and target graph, and the deformed graph template is used to build the mesh. Taking advantage of end-to-end learning and discarding direct dependency on annotated labels, our \ourmodel{} demonstrates outstanding performance in generating structured vascular meshes on cardiac and cerebral vascular images. It shows great potential for clinical applications by reducing mesh generation time from 2 hours (manual) to 30 seconds (automatic).


\end{abstract}
\section{Introduction}

Vascular diseases, such as heart attack and stroke, can affect the quality of human life and even cause disability and death.
Increasing attention has been paid to investigating the relationship between vascular pathology and geometry \cite{samady2011coronary,brown2016role}.
Patient-specific vessel analysis models derived from vascular images have been developed to simulate various aspects of vascular function, including hemodynamics, tissue mechanics, and fluid-structure interaction, to improve the performance of cerebral-cardiovascular diagnosis and treatment \cite{carpenter2020review, gholipour2018nonlinear, gholipour2018three}.
However, generating a reliable simulation model from vascular images is time-consuming, which limits its employment in clinics and population studies \cite{gholipour2018three}.

Computational dynamics simulation of vessels requires accurate geometries from vascular images \cite{gholipour2018three, carpenter2020review}. 
Besides, the dynamics simulation of coronaries requires the deformation of the coronary from a sequence of image snapshots of the coronary throughout a cardiac cycle, which requires adjustments of various parameters to achieve a high-quality mesh. 
When simulating dynamics with mesh models using the finite element method, unstructured elements are widely used, but they lead to higher computational costs and less accurate results than sweeping-based structured hexahedral elements \cite{vinchurkar2008evaluation, ghaffari2017large}.

Vascular segmentation from images is a commonly used manner to obtain vasculature. 
Although deep learning-based models~\cite{chen2022attention, qiu2023corsegrec,qi2023dynamic,zhang2023anatomy,wang2024avdnet} have shown superiority over traditional vessel segmentation algorithms,
existing segmentation models still have difficulty in obtaining satisfactory results at the pixel level due to the intrinsic characteristics in vascular images, such as tubular structures and extreme class imbalance problems\cite{jia2021learning,qiu2023corsegrec}.
Besides, even an acceptable vascular segmentation cannot guarantee a precise vascular topology (due to branching merging and disconnected vessels), and geometric smoothness (due to bulges or depressions on the vascular surface). Therefore, meshing based on vascular segmentation is inefficient and not straightforward.

In this paper, we utilize the concept of template deformation and generate hexahedral mesh through a graph convolutional network (GCN) directly from vascular images.
Our contributions can be summarized as follows:
    1) We propose an end-to-end mesh generation framework, named \ourmodel{}, which estimates the deformation of a centerline-based template graph to generate hexahedral vascular mesh from vascular images directly. 
    2) We obtain topological features, i.e., the tubular shape, of vascular images by a proposed sampling operator and employ a GCN to estimate the graph deformation. 
     3) We present a dedicated graph-based loss to estimate deformation of the template graph.
     4) We validate \ourmodel{} for the tasks of coronary and cerebral vascular meshing on multiple datasets. Experimental results show the effectiveness of our framework with significant improvement compared to the state-of-the-art methods.

\section{Method}
\label{method}

\begin{figure*}[t]
  \centering
  \subfigure[]{
  \includegraphics[width=0.6\linewidth]{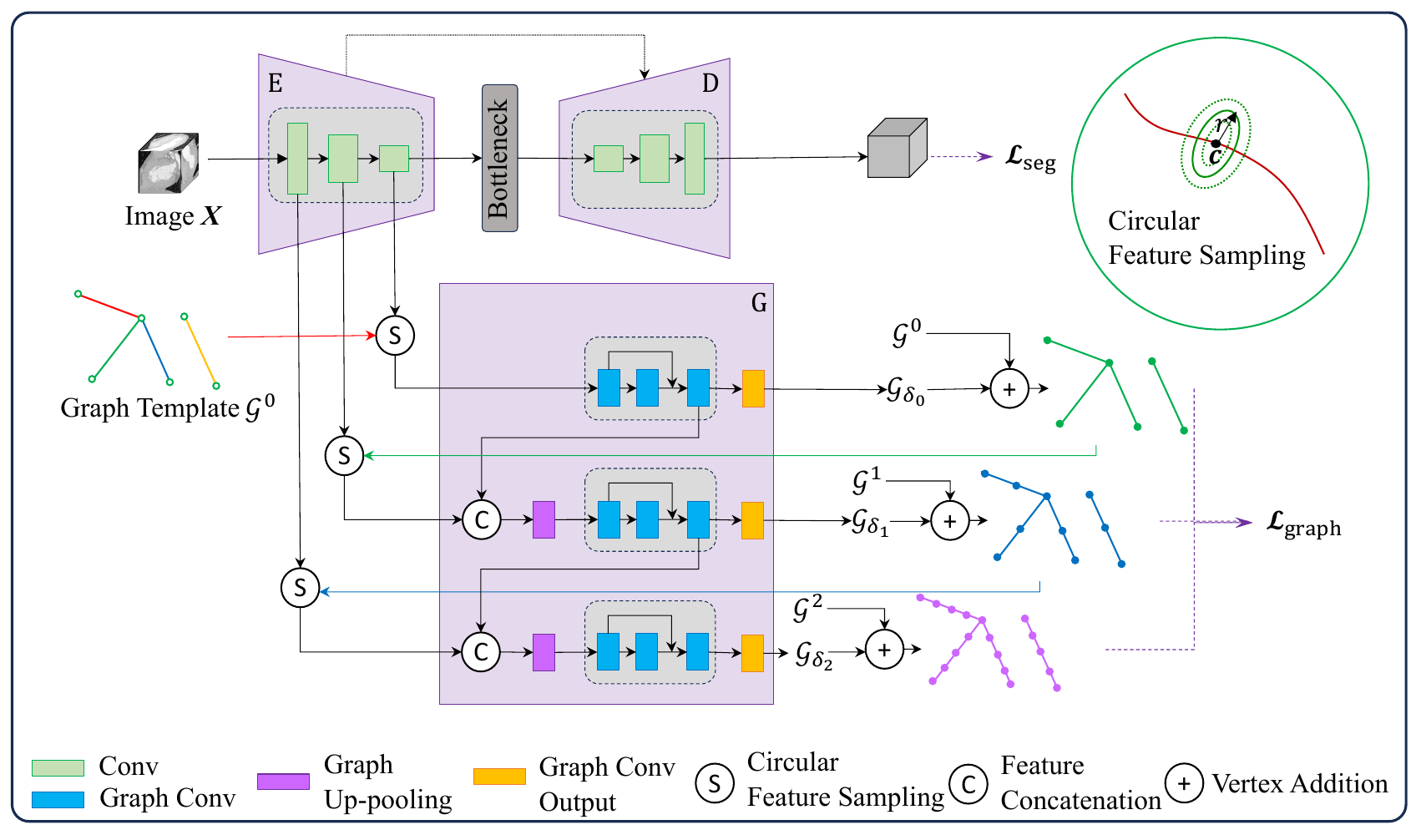}
  }
  \subfigure[]{
  \includegraphics[width=0.35\linewidth]{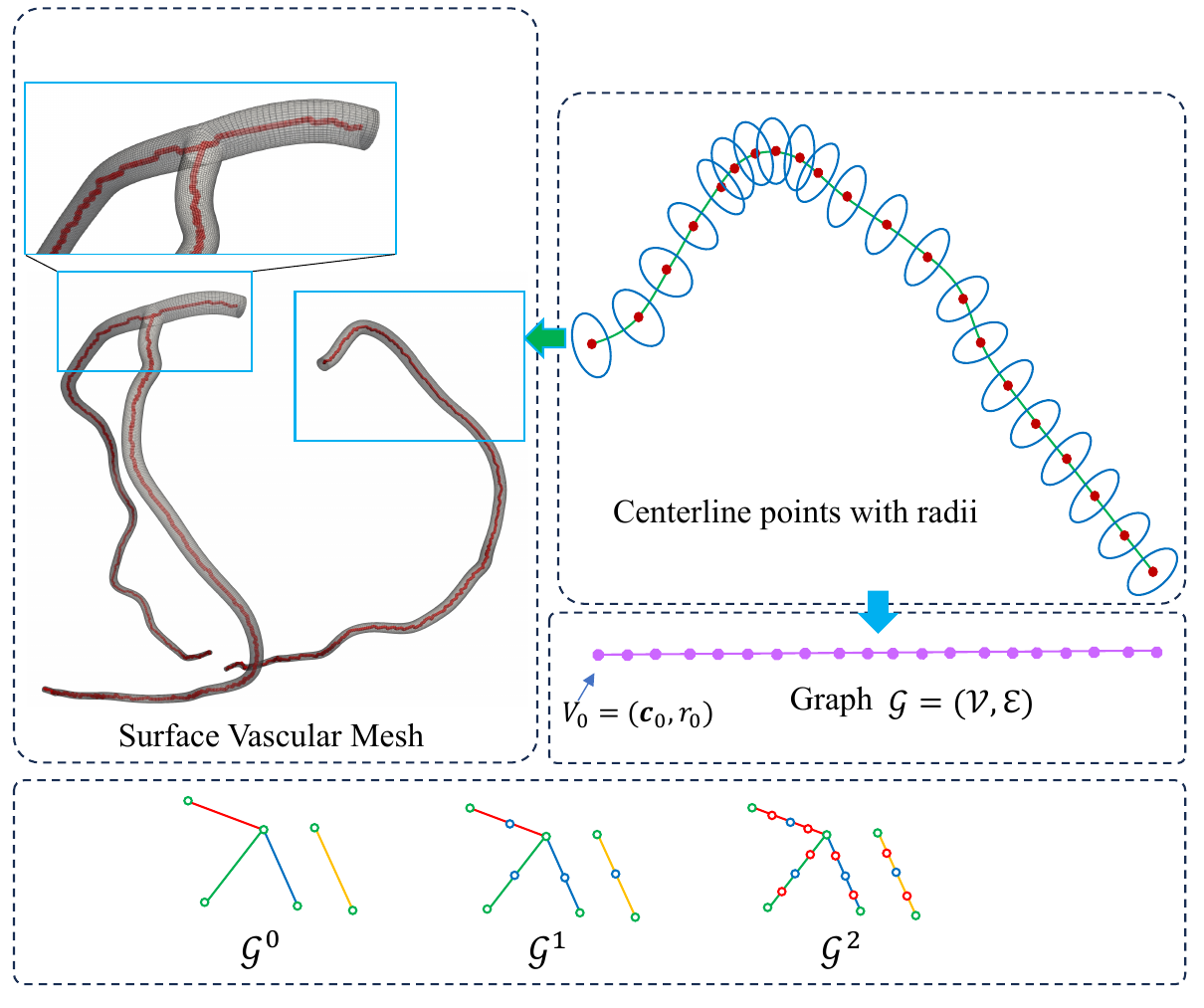}
  }
  \caption{Schematic illustration of our \ourmodel{}. (a) Network structure of our \ourmodel{}. Given a vascular image and a graph template, we sample the extracted features of the encoder of a segmentation network and embed the sampled features into a graph. Taking the graph as input, we use graph convolutional networks (GCN) to infer the graph deformation $\mathcal{G}_\delta$, such as deformations on vertices. More details are given in ``Method'' section. The subfigure in the green circle shows feature sampling centered at centerline point $\boldsymbol{c}$ with radius $r$ in the cross-section of a vertex. (b) A vascular surface mesh (left) and a graph (right). Here, a vertex $\mathcal{V}_0$ of the graph contains a four-element vector containing the 3D coordinates and radius of the cross-section, i.e., $\mathcal{V}_0=(x_0, y_0, z_0, r_0)$. The estimated graph is used to create the hexahedral vascular mesh by a sweeping technique. The bottom subfigure illustrates three graph templates from sparse to dense.
  }
  \label{fig:main}
\end{figure*}

Our model consists of a segmentation network and a graph convolutional network (GCN) as shown in Fig.~\ref{fig:main}(a). The segmentation network aims to extract the topological features of vascular images. These features are sampled based on vertices in the graph template by the proposed sampling operator. The sampled features construct a new graph, which is fed into a GCN. The output of GCN contains the template deformation, and the deformed template is used to construct the mesh. More details are given below.

\subsection{Definition and Construction of Graph Template}
\label{sec:gcn}
Let the vascular image be $\boldsymbol{X}$.
We define a graph template $\mathcal{G}^0 =(\mathcal{V}^0, \mathcal{E}^0)$, which consists of $m$ vertices $\mathcal{V}^0 =\{\mathcal{V}^0_i\}_{i=1}^m$, and $\mathcal{V}^0_i=(\boldsymbol{c}_i, r_i)=(x_i,y_i,z_i, r_i)$.
Here, $\boldsymbol{c}_i=(x_i, y_i, z_i)$ is the coordinates of the $i$th centerline point, and $r_i$ is the radius of the circular cross-section corresponding to point $i$.
The graph has a set of undirected edges $\mathcal{E}^0\subseteq \mathcal{V}^0 \times \mathcal{V}^0$, such that $(\mathcal{V}_{i_1}^0,\mathcal{V}_{i_2}^0)=(\mathcal{V}_{i_2}^0,\mathcal{V}_{i_1}^0)$.
Commonly, the edge $\mathcal{E}^0$ is described as a symmetric edge indicator matrix $\mathcal{E}^0\in \mathbb{R}^{m\times m}$, where $\mathcal{E}_{{i_1}{i_2}}^0= 1$ indicates a connection between $\mathcal{V}_{i_1}^0$ and $\mathcal{V}_{i_2}^0$; otherwise $\mathcal{E}_{{i_1}{i_2}}^0=0$. We construct our graph template referring to the method proposed in \cite{jia2019directional}.

\subsection{Feature Sampling for Multi-scale GCNs}
\label{sec:mesh}
To associate the vascular image with the template graph, we propose to sample the extracted features of the vascular image based on the vertices of the template graph. Specifically, as shown in Fig.~\ref{fig:main}(a), we employ a 3-stage U-shaped segmentation network
to extract topological features of the vascular image.
To locate the corresponding point of the vertex in the extracted features in each channel, we resize the vertex point coordinates to the feature size extracted by the segmentation network. 
As shown in the green circle (upper right corner) of Fig.~\ref{fig:main}(a), centered at the located corresponding point $\boldsymbol{c}$ with its radius $r$, we sample 48 points on each circle with the radii of $0.5r$, $r$ and $1.5r$.
All these sampled features are averaged as the feature of the current point. 
In such a way, the features of the vascular image are transformed into node features of the graph. 
We formulate this sampling operator as 
\begin{equation}
   \mathcal V^{\theta}_i =s(\mathcal V^0_i, e_{\theta}(\boldsymbol{X})),
   \label{eq:sample}
\end{equation}
where $s(\cdot)$ is the sampling operator and $e_{\theta}(\boldsymbol{X})$ represents the extracted features by the encoder, with $\theta$ being the network parameters. For each vertex of the graph template, we obtain the transformed features based on Eq.~\eqref{eq:sample}, and in conjunction with the edge $\mathcal E^0$ of the graph template, we yield the individual-specific graph $\mathcal{G}^{\theta} = (\mathcal V^{\theta}, \mathcal E^0)$. Subsequently, we forward this graph to a GCN, i.e., $g_{\phi}$, with learnable parameters $\phi$, to map the graph with image features into a new graph, i.e., $\mathcal{G}^{\theta,\phi}=g_{\phi}(\mathcal{G}^{\theta}) = (\mathcal V^{\theta,\phi}, \mathcal E^0)$. This new graph contains the deformation of the graph template.

In fact, to enable fine-grained sampling, we adopt a multi-scale scheme. Specifically, 
instead of using one GCN, we employ three GCNs for the three stages in the encoder, respectively. For the other two stages, we use graphs $\mathcal{G}^1$ and $\mathcal{G}^2$. $\mathcal{G}^1$ is generated by splitting each edge of graph $\mathcal{G}^0$ with a new vertex, resulting in two edges, and similarly, we split $\mathcal{G}^1$ to obtain $\mathcal{G}^2$. It is worth noting that all the graphs $\mathcal{G}^0$, $\mathcal{G}^\theta$, and $\mathcal{G}^{\theta, \phi}$ share the same edge information. In such a way, multi-scaled deformed graphs are estimated, i.e., $\mathcal{G}_{\delta_0}$, $\mathcal{G}_{\delta_1}$, and $\mathcal{G}_{\delta_2}$, from sparse to dense. Finally, we use the deformed graph with the finest grid as our output.

\subsection{Graph Loss for Estimating Template Deformation}
\label{graph_loss}
We construct our graph loss for learning template deformation in three levels. 
The first level is the subgraph level, which accounts for the branch group.
For example, there are two subgraphs of a coronary in the bottom subfigure of Fig.~\ref{fig:main}(b).
The second level is the vascular segment level.
For example, we have three vascular segments in the left subgraph of a coronary as shown in red, green, and blue colored lines shown in the bottom subfigure of Fig.~\ref{fig:main}(b).
The third level is the point level, which calculates the distance for each point as marked by small colored circles in the bottom subfigure of Fig.~\ref{fig:main}(b). 
Hence, our graph loss is given by
\begin{equation}\label{mesh_loss}
\mathcal{L}_{\rm{graph}}=\sum_{h=1}^{N_{\rm{sc}}}\sum_{n=1}^{N_{\rm{sg}}}\sum_{m=1}^{N_{\rm{vsm}}}\left(\sum_{p=1}^{N_{\rm{S}}}\mathcal{D}_{nmp}+\sum_{q=1}^{N_{\rm{T}}}\mathcal{D}_{nmq}+\lambda\mathcal{R}_{nm}\right),
\end{equation}
where  $N_{\rm{sc}}$, $N_{\rm{sg}}$ and $N_{\rm{vsm}}$ denote the numbers of scales, sub-graphs, and vascular segments, respectively. $N_{\rm{S}}$ and $N_{\rm{T}}$ indicate the numbers of points in the deformed template $\mathcal{G}^{\rm{S}}$ and target graph $\mathcal{G}^{\rm{T}}$, respectively. $\mathcal{R}_{nm}$ is the regularization term.

With respect to the definition of distance $\mathcal{D}$, since the number of points in the deformed template $\mathcal{G}^{\rm{S}}$ and target graph $\mathcal{G}^{\rm{T}}$ could be different, we define two symmetric losses, where each loss calculates the distance between a vertex of graph A and the nearest vertex in graph B. For instance, given a vertex $\mathcal{V}_{nmp}^{\rm{S}}=(\boldsymbol{c}^{\rm{S}}_{nmp},r^{\rm{S}}_{nmp})$, where $\mathcal{V}_{nmp}^{\rm{S}}$ denotes the vertex $p$ of vascular segment $m$ in subgraph $n$ of the deformed template.
The distance $\mathcal{D}_{nmp}$ in Eq.~\eqref{mesh_loss} is defined as $
 \mathcal{D}_{nmp}=\| \boldsymbol{c}^{\rm{T}}_*-\boldsymbol{c}^{\rm{S}}_{nmp}\|_2^2+( r_*^{\rm{T}}-r^{\rm{S}}_{nmp})^2,
$
where $
\boldsymbol{c}^{\rm{T}}_*=\mathrm{argmin}_{\boldsymbol{c}_i\in \boldsymbol{c}^{\rm{T}}}\| \boldsymbol{c}_i-\boldsymbol{c}^{\rm{S}}_{nmp}\|_2^2 $, and $r_*^{\rm{T}}$ is the corresponding radius of $\boldsymbol{c}^{\rm{T}}_*$.
Similarly, given a vertex $\mathcal{V}_{nmq}^{\rm{T}}=(\boldsymbol{c}^{\rm{T}}_{nmq},r^{\rm{T}}_{nmq}
) $ in the target graph, we have the distance $\mathcal{D}_{nmq}$ in Eq.~\eqref{mesh_loss} defined as $
 \mathcal{D}_{nmq}=\| \boldsymbol{c}^{\rm{S}}_*-\boldsymbol{c}^{\rm{T}}_{nmq}\|_2^2+( r_*^{\rm{S}}-r^{\rm{T}}_{nmq})^2,
$ 
where $
\boldsymbol{c}^{\rm{S}}_*=\mathrm{argmin}_{\boldsymbol{c}_i\in \boldsymbol{c}^{\rm{S}}}\| \boldsymbol{c}_i-\boldsymbol{c}^{\rm{T}}_{nmq}\|_2^2 $, and $r_*^{\rm{S}}$ is the corresponding radius of $\boldsymbol{c}_*^{\rm{S}}$.

To encourage a more uniform vertex density on the predictions, we define the regularization term in Eq.~\eqref{mesh_loss} as
\begin{equation}
\mathcal{R}_{nm}=\sum_{p=1}^{N_{\rm{S}}}\sum_{(\boldsymbol{c}_g^{\rm{S}},r_g^{\rm{S}})\in \mathcal{N}(\mathcal{V}^{\rm{S}}_{nmp})}\left|\| \boldsymbol{c}^{\rm{S}}_{nmp}-\boldsymbol{c}_g^{\rm{S}}\|_2^2-\alpha_{nm}^2\right|+\left|( r^{\rm{S}}_{nmp}-r_g^{\rm{S}})^2-\beta_{nm}^2\right|,
\end{equation}
where $\mathcal{N}(\mathcal{
V}_{nmp}^{\rm{S}})$ represents the neighborhood of the vertex $\mathcal{
V}_{nmp}^{\rm{S}}$.
$\alpha_{nm}$ and $\beta_{nm}$ denote the average differences of coordinates and radii of vascular segment $m$ in subgraph $n$, which are computed from the ground truth.
Since we have multi-scale graphs (Fig.~\ref{fig:main}(a)), we actually accumulate the $\mathcal{L}_{\rm{graph}}$ of all three graphs as our graph loss.

\subsection{Network Structure and Training Strategy}
\label{sec:network}
This section describes the architecture of GCN and the segmentation network, as well as the training strategy for estimating vascular template deformation.
As shown in Fig.~\ref{fig:main}(a), for the first deformation estimation stage, we adopt a residual GC module to map the feature-embedding graph into a deformed one. 
The GC module consists of three graph convolution blocks, and each block has a cascade of “G-Conv+ReLU”.
The output of this module has four channels. 
The first three are the deformations in 3D coordinates, and the last channel contains the difference in radius.
In the second and third stages, we concatenate the sampled image features with the output of the GC module of the previous stage.
After the concatenation, we use a graph up-pooling layer to obtain denser features, and the output layer will hence generate denser node deformations.

With respect to the segmentation network, the U-shaped network has three stages with channels 32-128-256. We use the Dice loss for the segmentation network.
Our \ourmodel{} is trained end-to-end by balancing the proposed graph loss as formulated in Eq.~\eqref{mesh_loss}, and the segmentation loss is notated as $\mathcal{L}_{\rm{seg}}=1-\frac{2\boldsymbol{Y}\cdot\boldsymbol{P}}{\|\boldsymbol{Y}\|_1+\|\boldsymbol{P}\|_1}$, where $\boldsymbol{Y}$ and $\boldsymbol{P}$
are the ground truth and predicted probability map, respectively.
Our total loss is formulated as 
\begin{equation}\label{total loss}
\mathop{\arg\min}_{\theta ,\phi }\mathcal{L}_{\rm{graph}} + \lambda_{\rm{seg}}\mathcal{L}_{\rm{seg}}.
\end{equation}

\subsection{Template Construction and Mesh Reconstruction}
In this work, we evaluate our model on both cardiac and cerebral
vascular datasets. 
Therefore, we build a coronary template using annotated centerlines and radii from 112 cases, and build a cerebral vascular template from 20 cases.
The process of template generation, as delineated in \cite{jia2019directional}, involves initially identifying the vessel most analogous to others, followed by a series of refinements to perfect the initial selection.
Each vascular segment of the template is built separately \cite{jia2019directional}.
In our experiments, the number of vertices of $\mathcal{G}^0$ is 33 (with 21 for left and 12 for right branch) for coronaries and 24 (with 12 for left and 12 for right branch) for cerebral vessels.

Our method prefers the hexahedral vascular mesh to the tetrahedral one because of the stable properties of structured hexahedral mesh \cite{carpenter2020review}.
As shown in Fig.~\ref{fig:main}(b), the hexahedral mesh of a 3D vessel can be generated from the graph by sweeping technology \cite{ghaffari2017large}.
In our experiments, for each vertex in the deformed graph template, a circular cross-section is divided into 48 points.
Using the technology in \cite{ghaffari2017large}, we can sweep the contours along the centerlines, and successive sections are connected to form a mesh surface (Fig.~\ref{fig:main}(b)).
Using a structured O-grid pattern based on the filled contours, we can also create the hexahedral volume mesh.
\section{Experiments and Results}
\label{sec:experiments&results}
\subsection{Datasets}
We use cardiac datasets and cerebral vascular datasets to validate the performance of our model on the task of vascular computing, i.e., centerline extraction, vessel segmentation, and mesh reconstruction.
\subsubsection{Cardiac datasets.}
Since coronary pathologies with serious consequences often occur in main branches \cite{collet2018left}, we select three main branches for all the coronaries according to the centerline annotations.
To validate the performance of our model, we use three datasets, including a public dataset from MICCAI 2020 Automated Segmentation of Coronary Arteries (ASOCA) challenge \cite{gharleghi2022automated}, an in-house computed tomography angiography (CTA) dataset, and an in-house time-series CTA dataset.
ASOCA provides 40 cardiac images with coronary annotations \cite{gharleghi2022automated}. 
Our first private coronary artery dataset (PCAD) includes 40 cardiac CTA images annotated with centerlines and lumen of the coronaries. 
Our second private cardiac CTA dataset (T-PCAD) contains two-time-series (i.e., diastolic and systolic phases) CTA from 20 patients.

We randomly split the 40 cases of ASOCA into two sets, 28 for training and 12 for testing. 
We perform the same split on PCAD.
Finally, we test the model, which is trained using PCAD, on the 20 cases of T-PCAD to show the potential applications of coronary motion tracking.

\subsubsection{Cerebral vascular datasets.}
Besides evaluation on coronaries, we also validate \ourmodel{} on a public cerebral vascular dataset, i.e., the IXI dataset \cite{ixi}. 
Since the original vascular annotations of the IXI dataset are noisy, we randomly select 60 cases and relabel two branches of the cerebral vessels by three experts, which is named SIXI dataset in our work.
SIXI are scanned using time-of-flight magnetic resonance angiography (TOF-MRA) sequence, and we use 20 cases for template construction, 28 cases for training, and the rest 12 for testing.

\subsection{Implementation Details}
\ourmodel{} is implemented by Pytorch, and optimized by the Adam optimizer with an initial learning rate of $10^{-5}$.
The learning rate is decreased every 10 epochs by a factor of 0.1, and the number of training epochs is set as 1000. 
All experiments were run on an NVIDIA RTX 3090 GPU equipped with 24G RAM.
The weighting parameter $\lambda$ of the regularization term is set as 0.01, and the segmentation weight $\lambda_{seg}$ is chosen as 1.

\subsection{Quantitative Evaluation}
\begin{table}[t]
\caption{Average vascular centerline overlap (OV in \%) \cite{jia2019directional} and label overlap in Dice (\%) using PCAD, ASOCA, and SIXI. Standard deviations are shown in parentheses. The best results are in \textbf{bold}.}
\setlength{\tabcolsep}{2.3mm}
\label{tab:overlap}
\resizebox{\linewidth}{!}{\begin{tabular}{c|ccc|ccc} 
\toprule
\multirow{2}{*}{Method} & \multicolumn{2}{c|}{PCAD}     & \multicolumn{2}{c|}{ASOCA} & \multicolumn{2}{c}{SIXI}   \\ 
\cline{2-7}
                        & \multicolumn{1}{c|}{OV}    & \multicolumn{1}{c|}{Dice}    & \multicolumn{1}{c|}{OV}    & \multicolumn{1}{c|}{Dice}   & \multicolumn{1}{c|}{OV}  & \multicolumn{1}{c}{Dice}    \\ 
\hline
FM+LS                
& \multicolumn{1}{c|}{$81.4\pm16.4$} & \multicolumn{1}{c|}{$71.8\pm10.9$} & \multicolumn{1}{c|}{$87.1\pm15.8$} & \multicolumn{1}{c|}{$69.4\pm9.5$} 
    & \multicolumn{1}{c|}{$88.1\pm12.6$}& \multicolumn{1}{c}{$72.7\pm5.6$}  \\

UNet+SK& \multicolumn{1}{c|}{$73.4\pm8.7$} & \multicolumn{1}{c|}{$75.5\pm6.2$} & \multicolumn{1}{c|}{$82.6\pm8.8$} & \multicolumn{1}{c|}{$83.1\pm6.7$} & \multicolumn{1}{c|}{$84.6\pm3.9$}& \multicolumn{1}{c}{$82.7\pm2.8$}\\ 
    
    Baseline& \multicolumn{1}{c|}{$75.5\pm7.9$} & \multicolumn{1}{c|}{$76.3\pm3.8$} & \multicolumn{1}{c|}{$83.1\pm7.5$} & \multicolumn{1}{c|}{$\boldsymbol{85.6\pm5.9}$} & \multicolumn{1}{c|}{$85.4\pm3.7$}& \multicolumn{1}{c}{$83.0\pm2.6$}\\ \hline
\ourmodel{} (Ours)                    
& \multicolumn{1}{c|}{$\boldsymbol{93.1\pm2.8}$} & \multicolumn{1}{c|}{$\boldsymbol{77.2\pm3.2}$} & \multicolumn{1}{c|}{$\boldsymbol{94.6\pm2.0}$} & \multicolumn{1}{c|}{$85.4\pm5.8$} & \multicolumn{1}{c|}{$\boldsymbol{94.5\pm1.7}$}& \multicolumn{1}{c}{$\boldsymbol{84.2\pm2.7}$}\\

\bottomrule
\end{tabular}}
\end{table}

\begin{table}[t]
\caption{Average distance (AI in $\rm{mm}$) \cite{jia2019directional} and Hausdorff distance (HD in $\rm{mm}$) using PCAD, ASOCA, and SIXI. Standard deviations are shown in parentheses. The best results are in \textbf{bold}.}
\setlength{\tabcolsep}{2.3mm}
\label{tab:distance}
\resizebox{\linewidth}{!}{\begin{tabular}{c|ccc|ccc} 
\toprule
\multirow{2}{*}{Method} & \multicolumn{2}{c|}{PCAD}     & \multicolumn{2}{c|}{ASOCA} & \multicolumn{2}{c}{SIXI}   \\ 
\cline{2-7}
& \multicolumn{1}{c|}{AI}    & \multicolumn{1}{c|}{HD}    & \multicolumn{1}{c|}{AI}    & \multicolumn{1}{c|}{HD}   & \multicolumn{1}{c|}{AI}  & \multicolumn{1}{c}{HD}    \\ 
\hline
    FM+LS& \multicolumn{1}{c|}{$0.38\pm0.19$} & \multicolumn{1}{c|}{$10.20\pm17.89$} & \multicolumn{1}{c|}{$0.32\pm0.37$} & \multicolumn{1}{c|}{$9.52\pm16.66$} 
    & \multicolumn{1}{c|}{$3.54\pm2.00$}& \multicolumn{1}{c}{$14.50\pm10.46$} \\
    
    UNet+SK& \multicolumn{1}{c|}{$0.40\pm0.17$} & \multicolumn{1}{c|}{$7.88\pm15.47$} & \multicolumn{1}{c|}{$0.34\pm0.31$} & \multicolumn{1}{c|}{$7.60\pm14.74$} & \multicolumn{1}{c|}{$2.57\pm1.56$}& \multicolumn{1}{c}{$9.58\pm8.43$}\\ 
    
    Baseline& \multicolumn{1}{c|}{$0.38\pm0.20$} & \multicolumn{1}{c|}{$7.26\pm13.49$} & \multicolumn{1}{c|}{$0.27\pm0.16$} & \multicolumn{1}{c|}{${6.11\pm15.15}$} & \multicolumn{1}{c|}{$2.41\pm1.48$}& \multicolumn{1}{c}{$9.04\pm7.30$}\\ 
    \hline
    \ourmodel{} (Ours) & \multicolumn{1}{c|}{$\boldsymbol{0.18\pm0.09}$} & \multicolumn{1}{c|}{$\boldsymbol{6.61\pm14.72}$} & \multicolumn{1}{c|}{$\boldsymbol{0.07\pm0.05}$} & \multicolumn{1}{c|}{$\boldsymbol{5.88\pm14.34}$} & \multicolumn{1}{c|}{$\boldsymbol{0.28\pm0.14}$}& \multicolumn{1}{c}{$\boldsymbol{8.62\pm4.51}$}\\
\bottomrule
\end{tabular}}
\end{table}

To quantitatively evaluate our \ourmodel{} in terms of centerline extraction and lumen segmentation, we compared our model with different representative ones. Specifically, we employed the fast-marching method to extract the centerlines \cite{jia2019directional} and adopted the level-set method \cite{li2010distance} to segment the lumen, denoted as FM+LS.
Besides, we trained the nnUNet \cite{isensee2021nnu} on the same dataset using the public source code and default settings. Based on the segmentation results, we used a skeleton approach \cite{lee1994building} to extract the centerlines, denoted as UNet+SK.
Moreover, we trained a baseline model, which shares the same architecture as the encoder-decoder network of \ourmodel{}, with its $\mathcal{L}_{\rm{seg}}$ in Eq. \eqref{total loss} and performed vascular segmentation.
The baseline model obtains centerlines using the skeleton approach.
For fair comparisons, all the learning-based models were trained using the same data augmentation scheme, including random flip, rotation, and elastic deformation.

Tables~\ref{tab:overlap} and ~\ref{tab:distance} present the results of compared methods in terms of overlap and distance. 
We can see that our \ourmodel{} achieves the best performance on centerline extraction task with significant improvement.
Notably, \ourmodel{} outperforms Baseline by a large margin.
It demonstrates that the proposed graph loss added to the GCN is the key factor to performance improvement.
In contrast, the conventional methods using a two-stage strategy, i.e., UNet+SK and Baseline, fall far behind our \ourmodel{} based on end-to-end learning.

\subsection{Qualitative Evaluation}

\begin{figure*}[t]
  \centering
 
  \includegraphics[width=0.7\linewidth]{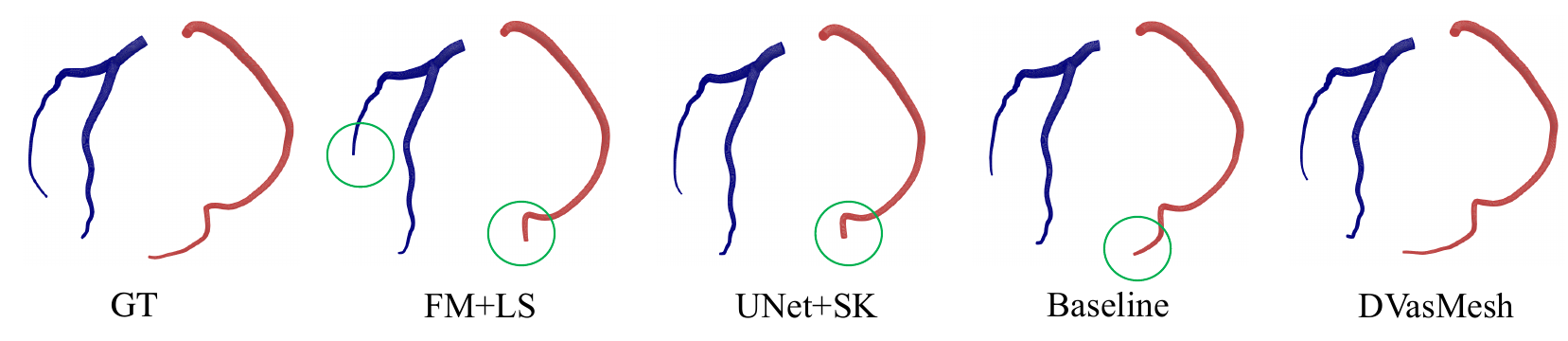}
  \caption{Qualitative comparisons of vascular volume meshes among different methods. The green circles mark the disconnected regions.}
  \label{fig:compare_results}
\end{figure*}

Fig.~\ref{fig:compare_results} illustrates the reconstructed mesh of coronary vascular volume by different methods.
We can see that our \ourmodel{} generates more anatomically consistent geometries compared to FM+LS and segmentation-based approaches (i.e., UNet+SK and Baseline), which encounter disconnection problems.

\begin{figure}[htbp]
\centering
\begin{minipage}[t]{0.5\textwidth}
\centering

\includegraphics[width=5.5cm]{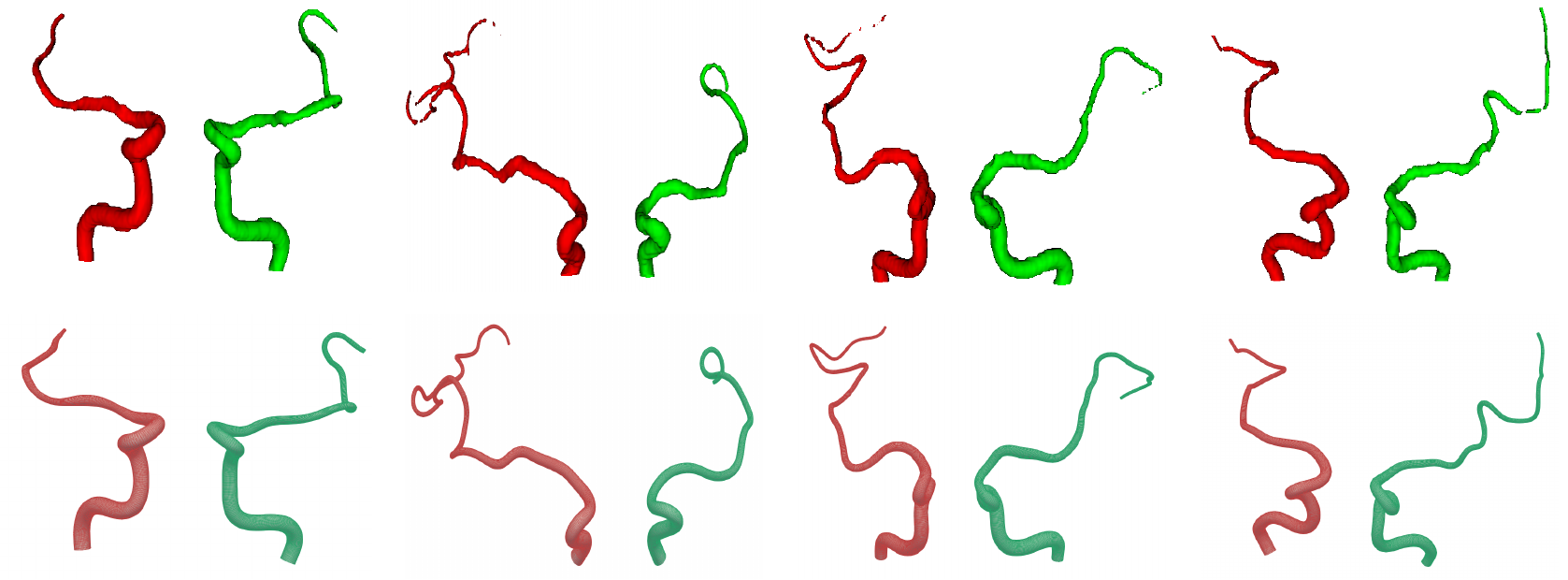}
\caption{Four examples of segmentation results (top) and vascular volume meshes (bottom) of left (red) and right (green) cerebral arteries reconstructed by \ourmodel{}.}
 \label{fig:cerebralvessel}
\end{minipage}
\begin{minipage}[t]{0.45\textwidth}
\centering
\includegraphics[width=4cm]{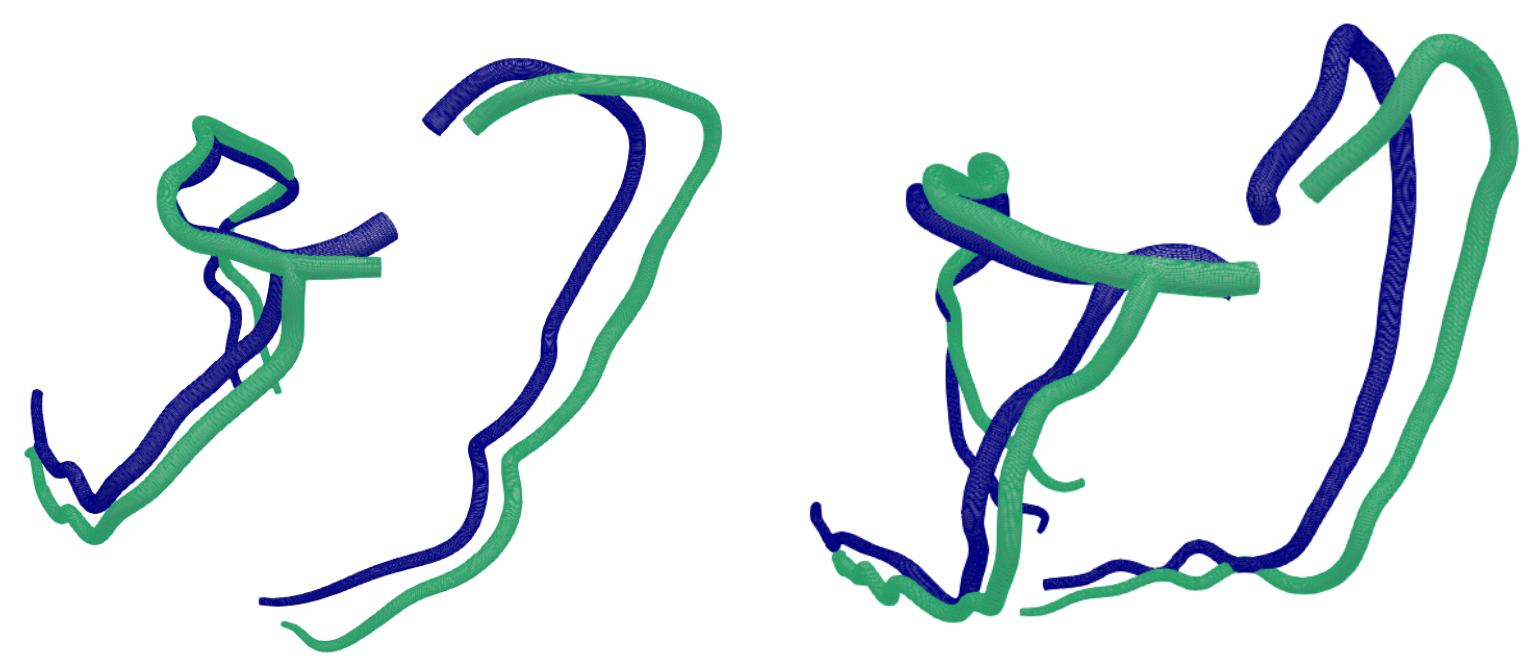}
\caption{Two examples of coronary volume meshes reconstructed from time-series cardiac CTA. The green and blue meshes are diastolic and systolic phases, respectively.}
 \label{fig:timeseries}
\end{minipage}
\end{figure}

Fig.~\ref{fig:cerebralvessel} shows vascular volume meshes of partial cerebral arteries.
We can observe that although the segmented vessels have disconnections in the distal parts, leading to undesirable geometries for the hemodynamic modeling of cerebral arteries,
our method is not affected by these disconnections, and reconstructs the volume meshes without discontinuity.

Moreover, we applied the trained model to the time-series cardiac CTA images, which consist of two-time frames over the cardiac cycle for each patient.
Fig.~\ref{fig:timeseries} shows the vascular meshes for two representative patients.
By deforming the same mesh template, we can track the position changes of the coronaries in diastole and systole.

\subsection{A Robust Solution to the Problem Vascular Disconnection}
A pressing issue of vascular segmentation and meshing is the vascular disconnection problem.
Many works attempt to reconnect the broken vessels, but carefully designed post-processing is still required to get improved results \cite{jia2019directional, qiu2023corsegrec}.
Our proposed \ourmodel{} is a template deformation approach based on deep learning, which can address this disconnection issue robustly.
The reconstruction accuracy \cite{qiu2023corsegrec} may differ for different cases using FM+LS method, but our method can consistently achieve perfect reconnection of the main branches of coronaries or cerebral vessels.
In the experiments, although we used partial branches of vessels, it is straightforward to extend to more branches when annotations are available.
As for diseased cases, the centerline, particularly the radii, can reflect pathological information, such as the severity of stenosis. A smaller radii value indicates more severe stenosis.

\section{Conclusion and Discussion}
In this work, we propose an end-to-end mesh reconstruction framework for vascular images. Different from the existing methods, we generate the vascular mesh by deforming a graph template, which is formulated using centerline points and radii. Besides, we propose a sampling scheme, which collects topological features of the vascular images based on the graph template. Furthermore, we employ these collected features to construct a graph, which is updated by graph convolution networks to estimate the graph deformation, and the deformed graph is followed by a sweeping technique to reconstruct the vascular mesh. 
Extensive experiments show that our model outperforms representative methods significantly for hexahedral mesh reconstruction of coronaries and cerebral vessels, and addresses the disconnection issue of vascular meshing.

The proposed method, rooted in template deformation, is intrinsically reliant on the structure of the template. 
In future endeavors, we will examine the impact of vascular templates with diverse architectures to facilitate comprehensive quantitative analyses.

\subsubsection{\ackname}
This work was supported fully by InnoHK Project (Project 2-3, H-CityU-28-32-3-PH2) at Hong Kong Centre for Cerebro-cardiovascular Health Engineering (COCHE).

\bibliographystyle{splncs04}
\bibliography{paper}

\end{document}